\documentclass[twocolumn,secnumarabic,amssymb, nobibnotes, aps, prl, superscriptaddress, nobalancelastpage,longbibliography]{revtex4-1}

 \usepackage{graphicx}
\usepackage{bm}% bold math
\usepackage{amsmath} \usepackage{braket} \usepackage{epsfig}
\usepackage{tensor}
\usepackage{CJKutf8}
\usepackage[version=4]{mhchem}
\usepackage{titlesec}
\usepackage{ifthen}
\usepackage{sidecap}
\usepackage{listings}
\usepackage[para,online,flushleft]{threeparttablex}
\usepackage[T1]{fontenc}

\usepackage{xr-hyper}
\usepackage{hyperref}
\hypersetup{breaklinks=true,colorlinks=true,linkcolor=blue,citecolor=blue,filecolor=magenta,urlcolor=cyan}

\usepackage[all]{hypcap}

\usepackage{xcolor}
\definecolor{pastelgray}{rgb}{0.81, 0.81, 0.77}
\definecolor{beaublue}{rgb}{0.9, 0.9, 0.93}

\makeatletter
\def\@bibdataout@aps{%
\immediate\write\@bibdataout{%
@CONTROL{%
apsrev41Control%
\longbibliography@sw{%
    ,author="08",editor="1",pages="1",title="0",year="1"%
    }{%
    ,author="08",editor="1",pages="1",title="",year="1"%
    }%
  }%
}%
\if@filesw \immediate \write \@auxout {\string \citation {apsrev41Control}}\fi
}
\makeatother

\renewcommand{\vec}[1]{\mbox{\boldmath $#1$}}

\begin{document}
\begin{CJK*}{UTF8}{gbsn}

\title{Nucleon-nucleon correlations in the extreme oxygen isotopes}

\author{S. M. Wang (王思敏)}
\affiliation{Institute of Modern Physics, Fudan University, Shanghai 200433, China}
\affiliation{FRIB Laboratory, Michigan State University, East Lansing, Michigan 48824, USA}

\author{W. Nazarewicz}
\affiliation{Department of Physics and Astronomy and FRIB Laboratory, Michigan State University, East Lansing, Michigan 48824, USA}

\author{R. J. Charity}
\affiliation{Departments of Chemistry and Physics, Washington University, St. Louis, MO 63130, USA}

\author{L. G. Sobotka}
\affiliation{Departments of Chemistry and Physics, Washington University, St. Louis, MO 63130, USA}

\date{\today}

\begin{abstract}
There has been an upsurge of interest in two-nucleon decays thanks to  the studies of nucleon-nucleon correlations. In our previous work, based on a novel time-dependent three-body approach,  we demonstrated that the  energy and angular correlations of the emitted nucleons can  shed  light on the structure of  nucleonic pairs formed inside the nucleus. In this work, we apply the new framework  to study the decay dynamics and properties of some extreme proton-rich and neutron-rich oxygen isotopes, including two-proton ($2p$) decays of $^{11,12}$O and two-neutron ($2n$) decay of $^{26}$O. Here we show that the low-$\ell$ components  of $^{11,12}$O wave functions, which are affected by  continuum and configuration-interaction effects,  strongly impact decay dynamics and asymptotic correlations.  In the calculated wave functions of 
 $^{11,12}$O,  diproton and cigarlike structures merge together during the tunneling process and the resulting energy- and angular correlations  are very consistent with the  experimental data.
The asymptotic correlations of the $2n$ decay of $^{26}$O dramatically change as the two-neutron decay energy  approaches the zero-energy threshold. The small reported value of  $Q_{2n}$ suggests that the $2n$ decay of this nucleus can be understood in terms of the universal phase-space limit.
\end{abstract}
\maketitle

\end{CJK*}

% Introduce pairing and correlation
{\it Introduction}.---Pairing is a ubiquitous feature of fermionic many-body systems that manifests itself in the phenomena of superfluidity and superconductivity  \cite{Bardeen1957,Cooper1959,Leggett2004}. Many low-energy properties of the atomic nucleus are profoundly affected by pairing between its proton and neutron constituents \cite{Brink2005,Broglia2013,Dean2003}. The pairing condensate present in nuclear ground states is responsible for the odd-even staggering of nuclear binding energies, which creates energetic conditions for the two-nucleon radioactivity observed in a handful of unbound rare isotopes \cite{Goldansky1960,Blank2008,Pfutzner2012}. However, the precise impact of pairing on the detected nucleonic pairs is largely unknown.  Numerous studies have been devoted to the question of  nucleon-nucleon correlations. For instance, the study of short-range correlations reveals the fundamental structure of nucleonic pairs at very high relative momenta  \cite{Hen2017}. However,  Cooper pairing that has profound impact on   ground-state (g.s.) properties of atomic nuclei, is a low-momentum phenomenon. The presence of pairing condensate in weakly-bound nuclei close to the particle dripline leaves a strong imprint on these systems \cite{Dobaczewski2007} as it impacts their binding, decay modes, and other properties. 
Since these systems are strongly affected by the presence of low-lying scattering channels, the pairing scattering into  the unbound (continuum) space is expected to  play a significant role \cite{Dobaczewski2013}.

% Previous work
In our previous work \cite{Wang2021}, we described the $2p$ decay  of $^6$Be and compared it with an artificial $2n$ decay of $^6$He using a recently developed time-dependent approach.  By comparing the dynamics of $2p$ emission with $2n$ decay, we showed that the  decay dynamics and long-range correlations are strongly impacted by  both initial-state and final-state interactions \cite{Watson1952,Migdal1955,Phillips1964}. In this Letter, we focus on heavier dripline systems, which allows us to gain more insight into the connection between inner structure and decay properties including  asymptotic correlations. 

% Oxygen isotopes
The recently discovered exotic oxygen isotope $^{11}$O, with 8 protons and only 3 neutrons,  has attracted significant attention due to its extreme proton-to-neutron ratio and interesting $2p$ decay characteristics \cite{Webb2019}. Its even-even neighbor, $^{12}$O, is also a $2p$ emitter \cite{Kryger1995,Jager2012}. Its  measured  energy and angular $2p$ correlations \cite{Webb2019_2}  are rather different from those observed in  $^6$Be \cite{Egorova2012}.  Another extreme oxygen isotope is the weakly neutron-unbound $^{26}$O, with 18 neutrons, which is arguably the best current candidate for the phenomenon of $2n$ radioactivity \cite{Kohley2015,Kondo2016,Grigorenko2013,Hagino2014,Adahchour2017,Fossez2017,Grigorenko2018,Li2021}. The measured $2n$ correlations in $^{26}$O have  large uncertainties, as well as its $2n$ decay energy $Q_{2n}$. However, the mere existence of this threshold resonance may have some unique consequences.

% The goal
In this Letter, we utilize our time-dependent approach to study the decay dynamics and resulting asymptotic correlations of $^{11,12,26}$O. By combining these results with our previous spectroscopic and structural studies  \cite{Wang2017,Wang2019}, we  aim at providing a comprehensive description of the extreme oxygen isotopes by revealing their open-quantum-system  nature.

{\it Method}.---In our three-body approach, the parent nucleus is described as a  core ($c$) plus two valence nucleons ($n_1,n_2$). The $i$-th cluster ($i=c,n_1,n_2$)  has the position vector $\vec{r}_i$ and linear momentum $\vec{k}_i$. The three-body Hamiltonian can be written as:
\begin{equation}\label{Hcnn}
	\hat{H} = \sum_i\frac{ \hat{\vec{p}}^2_i}{2 m_i} +\sum_{i>j} \hat{V}_{ij}(\vec{r}_{ij}) + \hat{H}_c -\hat{ T}_{\rm c.m.}.
\end{equation}
The second sum represents the pairwise interactions between the constituents, and $\hat{T}_{\rm c.m.}$ stands for the center-of-mass (c.m.) term. $\hat{H}_c$ is the core Hamiltonian given by the excitation energies of the core.

%Jacobi coordinates
To describe three-body asymptotics and to eliminate the spurious c.m. motion, one can define the Jacobi coordinates ($\vec{x},\vec{y}$) and relative momenta ($\vec{k}_x,\vec{k}_y$) \cite{Wang2021}. Noticing that there are two types of Jacobi coordinates ($Y$ and $T$ types), we use $\theta_k$ and $\theta_k^\prime$ to denote the opening angles of ($\vec{k}_x,\vec{k}_y$) in $Y$- and $T$-Jacobi coordinates, respectively \cite{Wang2021}. The  kinetic energy of the relative  motion  of the emitted nucleons is  $E_{pp/nn}$, and $E_{{\rm core}-p/n}$ is the kinetic energy of the core-nucleon sub-system.

%Wave function
The total wave function can be written as $\Psi ^{J\pi} = \sum_{J_v \pi_v j_c \pi_c} \left[ \Phi ^{J_v\pi_v} \otimes \phi^{j_c\pi_c} \right]^{J\pi}$, where
$\Phi ^{J_v\pi_v}$ denotes the wave function of valence nucleons and
$\phi^{j_c\pi_c}$ is
the  core wave function. 
For the core states, we assume a simple rotational picture; this allows the pair of valence nucleons to couple to the excited states of the core in a non-adiabatic way \cite{Barmore2000}. The valence wave function $\Phi ^{J_v\pi_v}$  is expressed in Jacobi coordinates and expanded with hyperspherical harmonics. A supersymmetric transformation method \cite{Sparenberg1997} is adopted to deal with the antisymmetrization between core and valence nucleons. For simplicity, we only project out those spherical Pauli-forbidden states that correspond to the  orbitals occupied by the core nucleons.

%GCC
The Gamow coupled-channel (GCC) method \cite{Wang2018,Wang2019} is utilized to calculate the initial wave function $\Psi^{J\pi}$(t=0), in which the hyperradial part of $\Phi ^{J_v\pi_v}$ employs the Berggren basis \cite{Berggren1968,Michel2009,Wang2017}. As a result, configuration space is extended to the complex-momentum $\tilde{k}$-plane. In this way, the predicted resonance has  complex energy $\tilde{E}$. Its real part represents the mean energy of the resonance while the imaginary part corresponds to the decay width ($-\Gamma /2$). The use of Jacobi coordinates and Berggren basis allows treatment of the inner and asymptotic regions of the Schr\"{o}dinger equation on the same footing, and prevents the reflection of the wave function at the boundary.

%TD
The complex-momentum state $\Psi^{J\pi}_{\rm GCC}$(t=0)  obtained with the GCC method is subject to purely outgoing (decaying) boundary conditions. In order to study the dynamics and asymptotic correlations of two-nucleon decay, this state can be decomposed into real-momentum scattering states using the Fourier–Bessel series expansion. The resulting $\Psi^{J\pi}_{\rm TD}$(t=0) is a wave packet and can be propagated within a time-dependent (TD) framework. To control the numerical precision, the time evolution operator is expanded with Chebyshev polynomials \cite{Volya2009,Loh2001,Ikegami2002}. Since we limit the time evolution to real momentum space, the wave function can be extended to  very large distances. This also helps to restore the Hermitian property of the Hamiltonian matrix and the conservation of total density. Meanwhile, the configuration mixing is involved using the same hyperspherical framework in both GCC method and TD approach. Consequently, studying the evolution of the coupled configurations allows us to analyze how the wave function structure evolves during the decay process.

{\it Hamiltonian and model parameters}.---Following our previous work on oxygen isotopes \cite{Wang2017,Wang2019}, in which energy spectra and  decay widths of $^{11,12,26}$O states were studied, in this work, we focus on the decay dynamics and nucleon-nucleon correlations.  As $^{11,12}$O and $^{26}$O are located in very different regions of the nuclear chart, different parameters are used to describe them. For  $^{11,12}$O, the core ($^{9,10}$C) is taken as a  rotor, which reasonably reproduces the intruder state containing the large $s_{1/2}$ component and includes the g.s. band of the core \cite{Barmore2000} (with $j_c^\pi \le 11/2^-$ for $^9$C, and $j_c^\pi \le 4^+$ for $^{10}$C). The effective core-nucleon interaction has been taken in the form of a Woods-Saxon (WS) potential (with spin-orbit term) and a one-body Coulomb interaction with the same parameters and charge distribution as Ref.\,\cite{Wang2019}. For $^{26}$O, we assume that the  $^{24}$O core is spherical, and the WS parameters are taken from Ref.\,\cite{Wang2017}.

%Interaction
The interaction between the valence nucleons is represented by the finite-range Minnesota force with the original parameters of Ref.\,\cite{Thompson1977}, which was fitted to the phase shifts from scattering data in a large energy range and has been widely applied to study structural properties of atomic nuclei, such as binding energies and spectra.  For the valence protons, this interaction is augmented by  the two-body Coulomb force.

\begin{figure}[!htb]
\includegraphics[width=0.95\columnwidth]{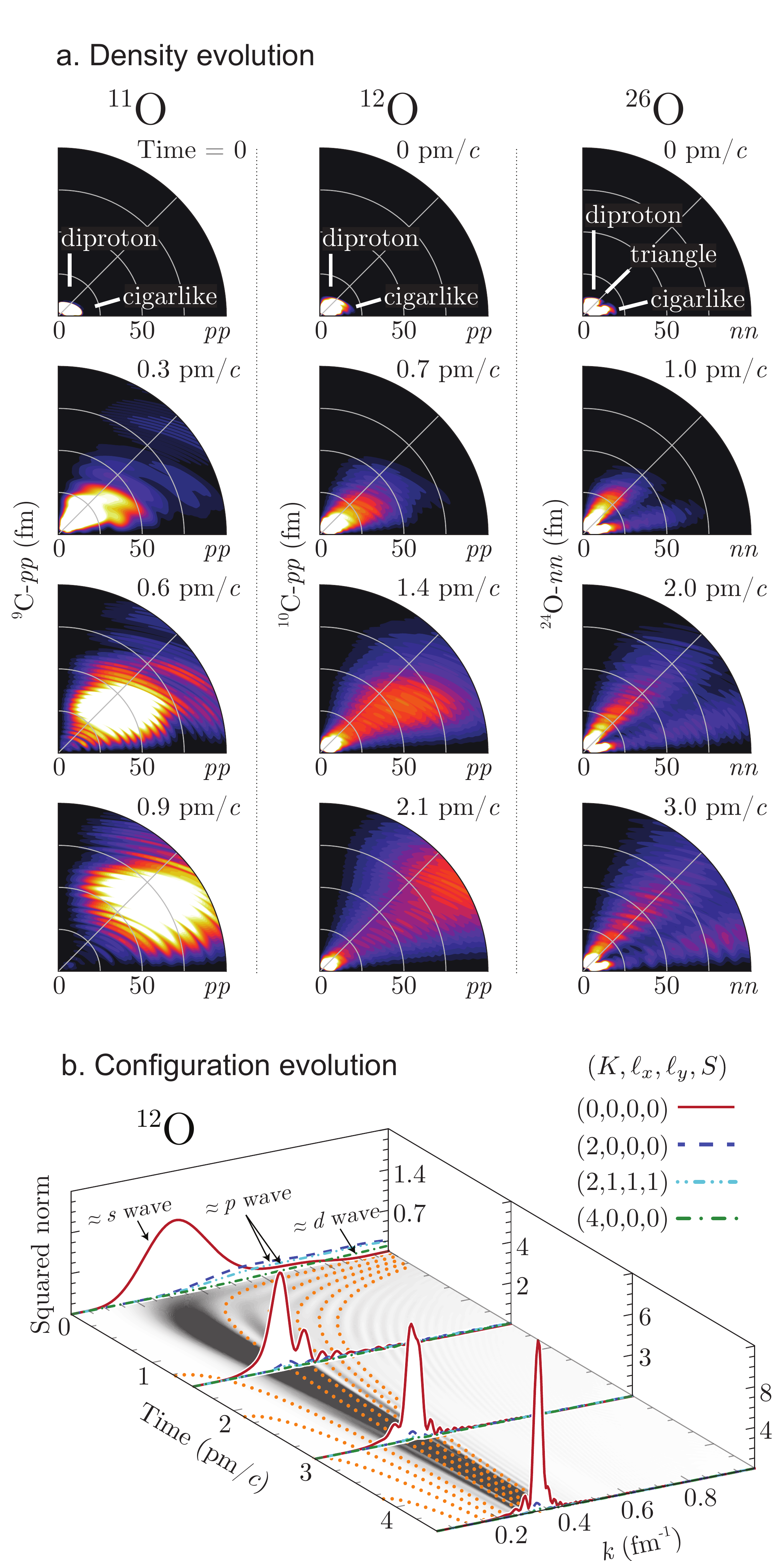}
\caption{(a) The density distributions of two-nucleon decays from the  g.s. of oxygen isotopes for four different time slices. (b) The  configuration evolution for $^{12}$O. The density distributions are shown in the Jacobi-$T$ coordinates, and multiplied by the polar Jacobi coordinate $\rho$ to highlight the asymptotic wave function. The Jacobi-$T$ configurations are labeled as $(K,\ell_x,\ell_y,S)$ in momentum  space. $k=\sqrt{{k_x^2}/{\mu_x} + {k_y^2}/{\mu_y}}$ is the total momentum  and ${\mu_x}$ (${\mu_y}$) is the reduced mass. The projected contour map represents the sum of all the configurations, in which the interference frequencies \cite{Wang2021} are marked by dotted lines.}\label{Time_evolution}
\end{figure}

%Model space
The three-body configurations in the Jacobi coordinates are denoted by $(K,\ell_x,\ell_y,S)$, where $K$ is the hyperspherical quantum number,  $\ell$ is the orbital angular momentum of the corresponding Jacobi axis, and $S$ is the total intrinsic-spin of the emitted proton (neutron) pair. Similar to Ref.\,\cite{Wang2019}, the calculations were carried out in a model space with $\max(\ell_{x}, \ell_{y})\le 7$ and for a maximal hyperspherical quantum number $K_{\rm max} = 20$. As shown in Ref.\,\cite{Wang2021}, the low-$\ell$ continuum is crucial during the decay process. Therefore, in the hyperradial part, we used the Berggren basis for the $K < 10$ channels and the harmonic oscillator basis with the oscillator length of 1.75\,fm and $N_{\rm max} = 40$ for the higher-angular-momentum channels. For the GCC calculation of the initial state, the complex-momentum contour defining the scattering part of the Berggren basis is given  by the path: $\tilde{k} = 0 \rightarrow 0.3-0.15i \rightarrow 0.5-0.12i  \rightarrow 1 \rightarrow 2 \rightarrow 4$ (all in fm$^{-1}$). For TD evolution, the inner part ($<15$\,fm) of the initial state is expanded and propagated with a real-momentum contour, which is $k = 0 \rightarrow 0.25 \rightarrow 0.5 \rightarrow 1 \rightarrow 2 \rightarrow 4$ (all in fm$^{-1}$). Each segment is discretized with 100  scattering states, which is sufficient to describe the outgoing wave function precisely. In practice, we only consider the interactions inside the sphere of radius 500\,fm. Since the wave function is  defined in the momentum space, and evolves from the highly localized initial wave packet, this cutoff  has no practical effect on the investigated physical observables.

%$2p$ decays of $^{11,12}$O
{\it 2${p}$ decay of proton-rich oxygen isotopes}.---The proton dripline  is located relatively close to the line of  $\beta$ stability. As a result, $2p$ correlation data have been obtained is several cases  \cite{Miernik2007,Webb2019_2}. In particular, the recently measured energy correlation of the emitted protons from the g.s. of $^{12}$O resembles that  of $^{16}$Ne (a 2$p$ emitter located in the $sd$-shell), but dramatically differs from correlations measured in    $p$-shell nuclei such as $^6$Be \cite{Webb2019_2}. This indicates there might be some structural similarity of  the configurations of the valence protons in $^{12}$O and $^{16}$Ne.

\begin{figure}[!htb]
\includegraphics[width=0.9\columnwidth]{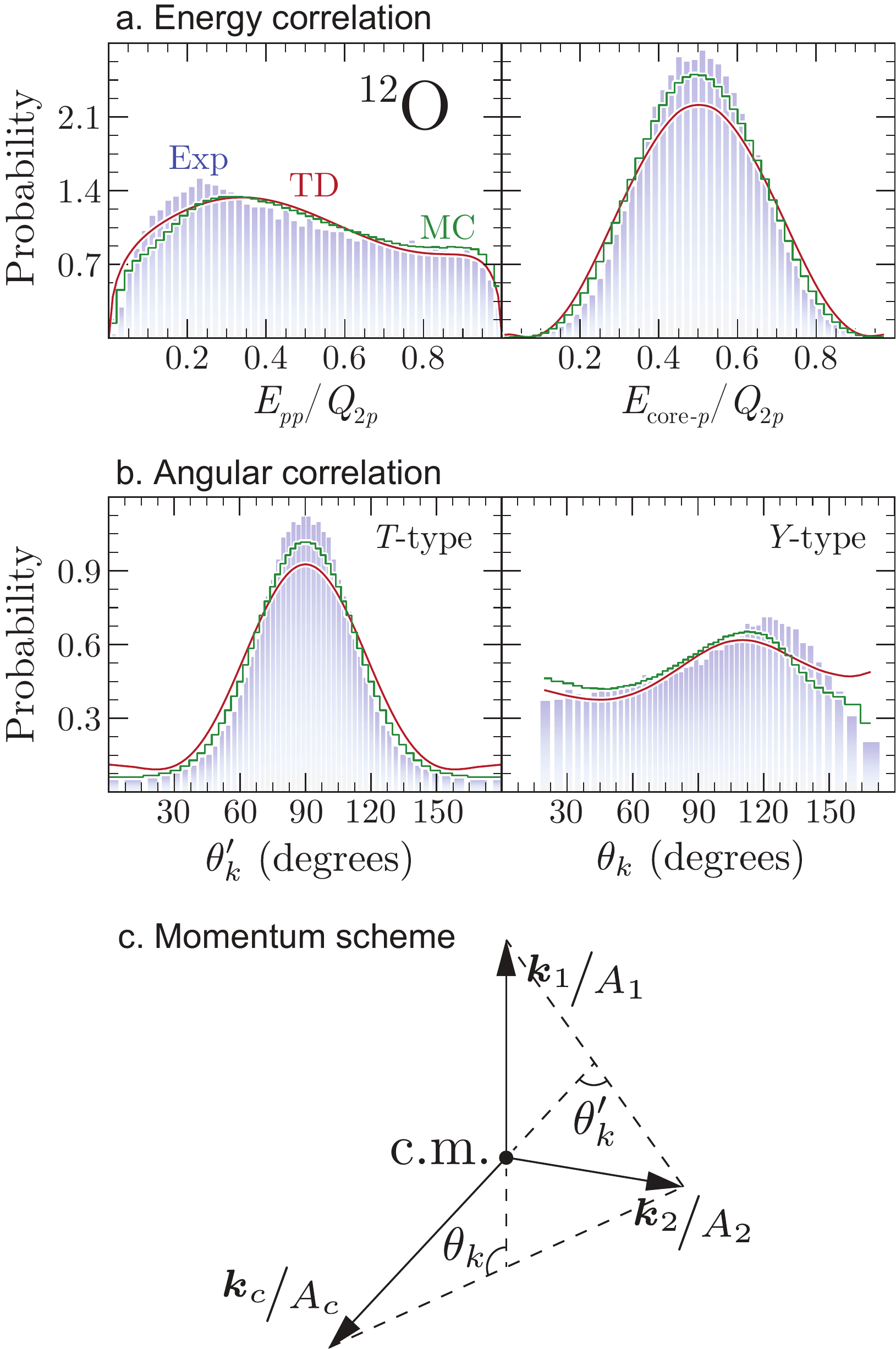}
\caption{Asymptotic (a) energy  and (b) angular correlations of the protons emitted from the two-proton unbound  $^{12}$O isotope. Also shown is (c) the momentum scheme for three-body system. Theoretical distributions were obtained within the time-dependent approach (TD) at ${t=15}$\,pm/${c}$. MC labels the Monte Carlo simulation of TD results which includes the experimental resolution and efficiency \cite{Webb2019_2}. The calculated $2p$ correlations (TD and MC) are compared with experimental data (Exp) of Ref. \cite{Webb2019_2}. $A$ is the mass number and $k_1$, $k_2$, and $k_c$ are the momenta of  the nucleons $n_1$ and $n_2$, and the core $c$, respectively in the c.m. coordinate frame.}\label{Correlation_O12}
\end{figure}

%Structure and decay dynamics of $^{12}$O
To understand the  $2p$ correlation patterns in $^{12}$O and  neighboring  $^{11}$O, we applied our time-dependent approach. The initial $2p$ density in $^{11,12}$O  is indicative of a pronounced diproton and a secondary cigarlike structure \cite{Webb2019,Wang2019}, which looks similar to that of the $p$-shell nuclei. However, for $^{12}$O, the emitted protons originating from these configurations merge together resulting in a broad distribution seen in Fig.~\ref{Time_evolution}(a), which shows dramatically different decay-dynamics compared to $^6$Be \cite{Wang2021}. This result is in agreement with the calculated flux current \cite{Wang2019}, which shows a competition between diproton and cigarlike decays constituting a democratic decay.

%Configuration of $^{12}$O
To gain more insight into this exotic decay dynamics, we can look into the details of those configurations. The   $S = 1$ component of the g.s. wave function of $^{12}$O, which has the squared amplitude of  27\% and allows for a more uniform distribution for the coordinate-space correlation inside the nucleus \cite{Wang2019}, is decimated asymptotically due to the finite orbital angular momentum components. On the other hand, the weight of Jacobi-$T$ configuration $(K,\ell_x,\ell_y,S)$ = (0,0,0,0) in $^{12}$O, approximately corresponding to the $s$-wave component, is dramatically enhanced (34\% for the initial state; see Fig.\,\ref{Time_evolution}(b)). A similar situation also occurs in $^{11}$O. This is due to the appearance of $s$-wave threshold resonances in the neighboring nuclei $^{10,11}$N \cite{Wang2019}. These poles of the scattering matrix in the complex momentum plane can be viewed as analogs of antibound (virtual) states in the mirror neutron-rich partners; their existence can be also beneficial for forming the diproton structure. Due to their the small centrifugal barriers, these low-$\ell$ components become dominant during the decay process. Figure\,\ref{Time_evolution}(b)  shows a transition in the  valence-proton wave function from a structure with moderate $p$- and $d$-wave components to one that is overwhelmingly  $\ell=0$ during the $2p$ decay of $^{12}$O.
Indeed, the weight of the  (0,0,0,0) configuration is  76\% in the final state. This $s$-wave enhancement is partly due to the coupling to the excited core states. 

%$2p$ correlation of $^{12}$O
The calculated asymptotic $2p$ correlations for $^{12}$O shown in Fig.\,\ref{Correlation_O12} are in qualitative, if not quantitative, agreement with experimental data. The minor differences between the experimental data and the Monte Carlo filtered calculations where the experimental acceptance and resolution is added to the time-dependent predictions could likely be reduced by modification of the employed original Minnesota force used for the nucleon-nucleon interaction. Another interesting aspect is that, unlike $^6$Be (see Fig 13(a) in \cite{Webb2019_2}), there is  a  low-energy peak  in the $E_{pp}$  correlation of $^{12}$O. As distinct diproton emission during the time evolution is not expected, a low-$E_{pp}$ correlation does not necessarily correspond to diproton decay. It is worth noting that the $2p$  system can form a subthreshold resonance with a broad decay width around 1\,MeV \cite{Kok1980}. This continuum feature is likely to affect  the energy correlation of the $2p$ emitters having small decay energies.

\begin{figure}[!htb]
\includegraphics[width=0.9\columnwidth]{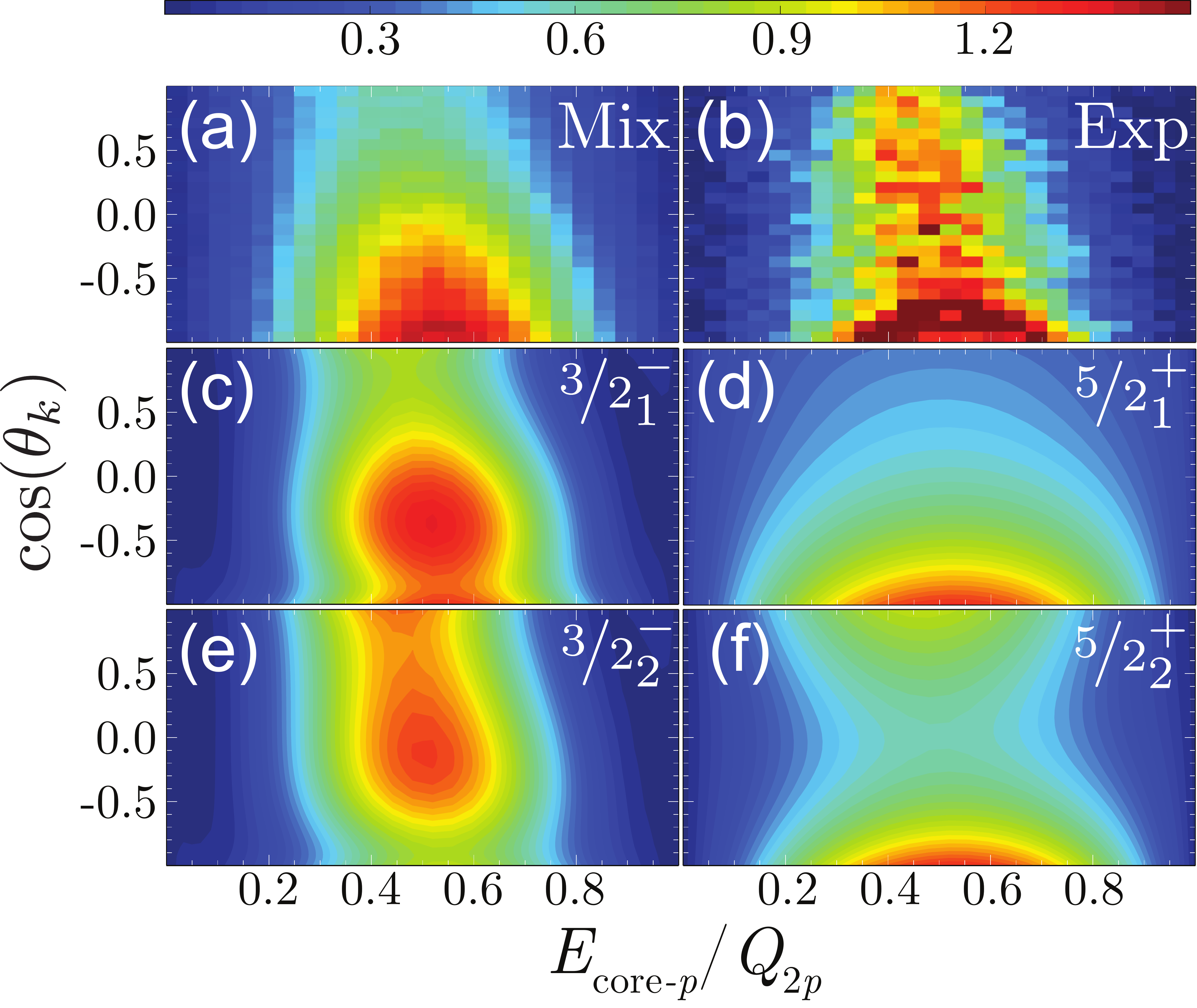}
\caption{(a) Theoretical and (b) experimental Jacobi-$Y$ correlations of two protons emitted from the broad low-energy  structure in $^{11}$O, and (c-f) the corresponding contributions from each low-lying state predicted. The experimental resolution and efficiency have been taken into account in (a) through Monte Carlo simulations.}\label{Correlation_O11}
\end{figure}

%$2p$ correlation of $^{11}$O
The lightest oxygen isotope $^{11}$O has been observed  as a  broad structure \cite{Webb2019} containing multiple resonances  \cite{Webb2020,Fortune2019,Garrido2020,Mao2020}. In our previous work, four low-lying states ($J^\pi$ = $3/2^-_1$, $5/2^+_1$, $3/2^-_2$, and $5/2^+_2$) were predicted in the experimental energy interval \cite{Wang2019}, each having a large decay width from 1 to 2\,MeV. In our TD calculations, these states have been propagated individually, i.e.,  possible interference effects have been neglected. The large decay widths  result in a strong continuum coupling, and a more uniform density distribution during the decay process compared to  $^{12}$O, see Fig.\,\ref{Time_evolution}. The resulting $Y$-type correlations show a strong dependence on the angular momentum, which could be useful to determine experimental spin assignments. To predict the asymptotic correlations of the valence protons emitted, we combine the correlations of the four low-lying states with the weights obtained by the resonance-shape fitting \cite{Webb2019}. This mixed correlation, shown in Fig.\,\ref{Correlation_O11}(a), captures the observed experimental features, Fig.\,\ref{Correlation_O11}(b). This agreement supports the argument that the observed broad structure is a mixture of $J^\pi$ = $3/2^-$ and $5/2^+$ states \cite{Webb2019, Webb2020}.

\begin{figure}[!htb]
\includegraphics[width=1\columnwidth]{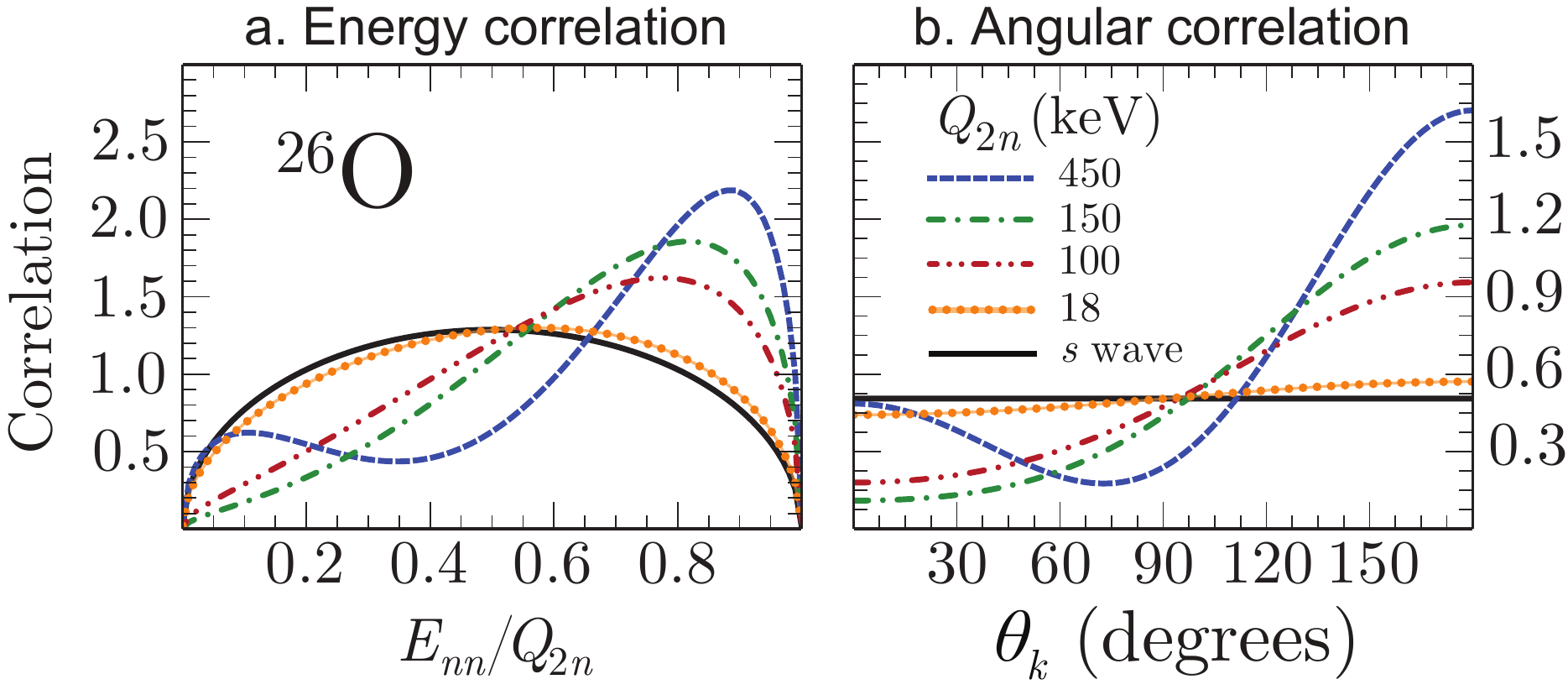}
\caption{ Asymptotic energy (a) and angular (b) correlations of emitted neutrons from the g.s. of $^{26}$O for different $2n$ decay energies $Q_{2n}$. Also shown are the analytical results for the asymptotic correlation in the  limit of $Q_{2n}=0$ (zero-energy $s$-wave). $\theta_k$ is the opening angle in the Jacobi-$Y$ coordinate, and $E_{nn}$ is the kinetic energy of the relative motion of the emitted neutrons.
}\label{Correlation_O26}
\end{figure}

{\it Two-neutron decay of threshold resonance in $^{26}$O}.---On the neutron-rich extreme, $^{26}$O is expected to decay via $2n$ emission \cite{Kohley2015,Kondo2016,Grigorenko2013,Hagino2014,Hagino2016,Grigorenko2018}.
According to our model,  $^{26}$O shows a very different structure, decay dynamics, and nucleon-nucleon correlation pattern as compared to the $2p$ emission from $^{11,12}$O (see Refs.\,\cite{Wang2017,Wang2019} and Figs.\,\ref{Time_evolution},\,\ref{Correlation_O12},\,\ref{Correlation_O26}). 

%$2n$ decay of $^{26}$O
Besides a dineutron and cigarlike structures,  the initial  $2n$ density of $^{26}$O also contains a triangular structure. These three configurations  are characteristic of the $d$-wave component. The very small $Q_{2n}=18\pm 5$\,keV value \cite{Kondo2016} in $^{26}$O  makes this nucleus a candidate for the $2n$ radioactivity.  When approaching the threshold, the presence of  the centrifugal barrier is expected to give rise to  changes in asymptotic correlations. As seen in Fig.\,\ref{Time_evolution} two branches emitted from the internal region are predicted, mainly corresponding to dineutron and large-angle configurations. The valence neutrons  maintain their correlation after tunneling, which shows an enhanced large-$E_{nn}$ correlation in Fig.\,\ref{Correlation_O26}. However, as we readjust the depth of the WS potential to bring the system closer to the experimental threshold, both energy and angular correlations of $^{26}$O become almost uniformly distributed, see Fig.\,\ref{Correlation_O26}. As discussed in Ref. \cite{Grigorenko2018}, for small values of  $Q_{2n}$ the $2n$ decay is dominated by the $s$-wave component and  the energy distribution approaches the universal phase-space limit:
\begin{equation}\label{Enn}
 \frac{d\sigma}{d\varepsilon} \sim \sqrt{\varepsilon(1-\varepsilon)}~~{\rm with}~~\varepsilon=E_{nn}/Q_{2n}.  
\end{equation}
At the same time, the angular distribution becomes essentially isotropic. This asymptotic behavior is practically reached in our calculations for the experimental value of $Q_{2n}=18$\,keV. As the energy of the resonance increases, asymptotic energy and angular correlations quickly start deviating from the phase-space limit (\ref{Enn}). It is seen that this deviation is already noticeable at $Q_{2n}=100$\,keV.

The appearance of $s$-wave dominated structures just above the reaction  threshold is well known \cite{Wigner1948,Barker1964}. When this happens, the threshold state becomes structurally aligned with the threshold and this leads to clustering effects  \cite{Okolowicz2020}. For $^{26}$O, the valence neutron pair is expected to be well decoupled from the core nucleus $^{24}$O due to the very small Q$_{2n}$ value.

{\it Summary}.---The main objective  of this Letter is to investigate the connection between the inner structure of the atomic nucleus  and the asymptotic correlations seen in two-nucleon decays. To this end, we studied two-proton decays of $^{11,12}$O and  two-neutron decay of $^{26}$O. Through time-dependent simulations, we have demonstrated that the structure of the initial wave function, governed by the initial-state  and final-state  interactions can impact the decay dynamics and leave an imprint on asymptotic correlations.

Due to the configuration mixing  and continuum coupling, initial wave functions of $^{11,12}$O contain a large $s$-wave component. In addition,  the considerable $S = 1$ part of the wave function  makes the coordinate-space correlations more uniformly distributed inside the nuclei. As a result, diproton and cigarlike structures merge together during the tunneling process. These initial configurations  manifest themselves in the asymptotic correlations, which show distinct patterns compared to the previously investigated case of case of $^6$Be \cite{Wang2021}. These patterns are consistent with the experimental correlation data for $^{11,12}$O.

In $^{26}$O, the threshold effect  dramatically changes the decay mechanism and asymptotic correlations. 
According to our calculations, the small reported value of  $Q_{2n}=18\pm 5$\,keV suggests that the $2n$ decay of this nucleus can be understood in terms of the universal $s$-wave phase-space limit \cite{Grigorenko2018}, in which the g.s. of $^{26}$O attains the character of  the core nucleus $^{24}$O to which a pair of neutrons is weakly coupled \cite{Okolowicz2020}.

{\it Acknowledgements}.---Discussions with Marek P{\l}oszajczak are gratefully acknowledged. This material is based upon work supported by the U.S.\ Department of Energy, Office of Science, Office of Nuclear Physics under award numbers DE-SC0013365 (Michigan State University), DE-SC0018083 (NUCLEI SciDAC-4 collaboration), and DE-FG02-87ER-40316.

\bibliography{references}

\end{document}